# TRANFORMATIONS INDUCED BY HYDROSTATIC PRESSURE ON LEAD METASILICATE PHASES

Ariano D. Rodrigues[1,*], Thiago R. Cunha[2], Rafaella B. Pena[3], Ulisses F. Kaneko[4], Lucas M. E. Pinho[1], Benjamim J. A. Moulton[5], Paulo S. Pizani[1]

[1]*Federal University of São Carlos (UFSCar), Physics Department, 13565-905, São Carlos, Brazil.*

[2]*Institute of Technology and Research (ITP), Center for Studies on Colloidal Systems, 49032-490, Aracaju, Brazil.*

[3]*University of São Paulo (USP), São Carlos Institute of Physics, 13566-590, São Carlos, Brazil*

[4]*São Paulo State University (UNESP), Institute of Geosciences and Exacts Sciences, 13506-900, Rio Claro, Brazil.*

[5] *Alfred University, 14802, Alfred, NY, United States*

## Abstract

Lead silicates are particularly notable within the silicate family due to their ability to form glass and glass-ceramics over an extended range of compositions, which enables a great variety of applications ranging from domestic to high technology. For most of them, controlling the crystallization – through the nucleation, growth, and stabilization of distinct crystalline phases – is critical to achieving the desired physical properties in the final glass-ceramic product. In this context, lead metasilicate $PbSiO_3$ (PS) represents an ideal model system for investigating structural evolution under varying pressure and temperature conditions. This is primarily due to its distinct Raman signatures and the capability of resolving its structure with high precision through diffraction measurements. These attributes enable a comprehensive evaluation of the thermodynamic quantities involved in this complex process, which are essential for the physical description of the crystallization of glasses undergoing heterogeneous nucleation. We report on high-pressure in situ analyses of three crystalline phases of PS: a stable monoclinic structure, a metastable hexagonal structure, and a lower symmetry metastable phase. Combined high-pressure Raman and synchrotron X-ray diffraction indicate that the structures are highly sensitive to the application of hydrostatic pressure and that significant structural rearrangements can be achieved in moderate pressure regimes (< 6GPa). Such analyses also enabled determining important thermodynamic variables of those systems, such as compressibility. From an applied perspective, our findings demonstrate that the application of pressure achievable using large-volume presses and capable of altering the energy states of such phases, can be regarded as a promising strategy to influence the stages of the overall crystallization process. This approach opens new avenues for the development of novel structures and properties in the resulting glass-ceramic materials.

---

[*] *Corresponding author* (ariano@df.ufscar.br)

## 1. Introduction

The crystallization processes of many silicate glasses, which give rise to glass-ceramic materials omnipresent from domestic use to high technology[1-6], do not follow a straightforward path. Commonly driven by high-temperature treatments, these processes involve multiple stages, during which the systems transition through intermediate metastable phases before attaining their more stable final conformation, indicating adherence to a form of Ostwald's rule of stages[7]. In the particular case of lead metasilicate ($PbSiO_3$, PS), existing literature has pointed out that temperature-induced crystallization – under ambient pressure – is a complex process involving metastable structures throughout its progression[8-10]. Beyond its relevance to technological applications, including infrared laser production[11], photonic crystals for telecommunications[12], and radiation shielding[13], the distinct Raman signatures and diffraction peaks associated with each crystalline polymorph establish PS as a model system for studying the mechanisms of multi-stage crystallization. Motivated by this, we have recently developed protocols optimized to stabilize, through the overall crystallization, the metastable and stable phases of PS [14], which under specific temperature treatments, enable to tune the glass-ceramics microstructure. At room conditions, PS glass presents three distinct crystalline polymorphs, the stable alamosite phase with a monoclinic structure (A-PS) and two metastable phases, one with hexagonal symmetry (H-PS) and the other with lower symmetry (L-PS). Note that the 'stable' and 'metastable' nomenclatures here refer to the stability against temperature, as presented in the related literature. Particularly noteworthy is the ability to perform in situ analyses as a function of pressure which, in addition to provide access to thermodynamic information for each phase, can potentially induce significant changes in local and group symmetries and energy landscapes. This approach allows for the investigation of thermodynamic conditions that can modify the multi-stage crystallization process, deviating from the conventional temperature-driven thermodynamic pathway and offering a promising strategy for producing glass-ceramic materials with unique properties. Furthermore, the broader significance of the knowledge gained from the studies reported below lies in its potential for transferring to other silicate compositions that also exhibit heterogeneous nucleation, thereby revealing its novel applicability.

## 2. Experimental

A bulk $PbSiO_3$ (PS) glass was synthesized using the standard splat-quenching method, with high-purity $SiO_2$ and $Pb_3O_4$ as precursors. Departing from the lead metasilicate glass, the three crystalline PS phases, namely A-PS, H-PS and L-PS, were individually obtained following the synthesis conditions reported elsewhere[10]. Quasi-hydrostatic high pressure was applied to the samples using a diamond anvil cell (DAC) with helium gas as the pressure-transmitting medium. The pressure within the DAC was determined via the standard ruby fluorescence method [15, 16], using a ruby sphere loaded alongside the sample. High-pressure Raman scattering measurements were performed using a 532 nm wavelength laser as the excitation source and recorded with a Horiba LabRam HR800 spectrometer (Physics Department, UFSCar) equipped with ultra-low-frequency filters, enabling the detection of wavenumbers as low as 10 $cm^{-1}$. The experimental setup was configured to achieve a spectral resolution of approximately 0.7 $cm^{-1}$, and these spectral parameters were consistently applied to all Raman measurements presented in this study. High-pressure synchrotron X-ray diffraction (HP-XRD) data were collected at the EMA beamline (Extreme Condition Methods of Analysis) at Sirius/CNPEM, using membrane-controlled DACs coupled to the X-ray micro focused beam with energy of ~ 25 keV ($\lambda$ = 0.0496nm) and diameter of 2 $\mu m^2$ [17]. The two-dimensional XRD patterns were recorded using a MARCCD X-ray photon detector, and the corresponding intensity versus 2θ XRD profiles were generated using the Dioptas software. The pressure was monitored through the luminescence of a ruby sphere loaded in the DAC, excited by a 532 nm wavelength laser and recorded by Princeton spectrometer HRS300 integrated with the beamline.

## 3. Results and discussions

### *3.1. Alamosite phase (A-PS)*

Figure 1a) highlights the two most significant regions of the Raman spectrum for the alamosite $PbSiO_3$ (A-PS) phase as a function of high pressure. In the high-wavenumber range (800 - 1150 $cm^{-1}$), the vibrations are associated with the fundamental building blocks, the $SiO_4$ units. In silicates, each $SiO_4$ tetrahedron can be corner connected up to four adjacent tetrahedral units, with each configuration corresponding to a specific vibrational mode. These modes contribute with distinct bands to the Raman spectrum at characteristic wavenumbers [18-22]. Variations in these peaks under pressure, such as shifts or changes in relative intensities, are indicative of structural rearrangements within the silicate environment, including alterations in bonding distances or tetrahedral connectivity [23-28]. The lower-wavenumber spectral region, specifically below 200 $cm^{-1}$, is dominated by the Pb-O vibrations, which present reduced vibrational frequencies due to the substantial atomic mass of lead [29-34]. As pressure increases up to approximately 29 GPa, the peaks shift toward higher wavenumbers, indicating the hardening of vibrational modes. Additionally, abrupt changes in the Raman spectra are observed at specific pressure values. These spectral variations, marked by transitions from spectra represented in red to black and subsequently to blue, delineate three distinct pressure regimes. Abrupt deviations in the eigenvalues of vibrational modes from the continuous stiffening trend with increasing pressure are typically indicative of significant structural transitions comprising changes in the local symmetry. In addition to peak shifts, the Raman profiles are also modified by changes in the relative intensities of peaks, particularly at pressures exceeding 13 GPa. The decrease in the intensities of high-wavenumber Raman bands (800 – 1150 $cm^{-1}$) relative to those in the lower-wavenumber range has been attributed to distortions in the $SiO_4$ units, preserving the tetrahedral coordination[35]. A quantitative analysis of the spectral changes within the higher-wavenumber range was conducted through spectral fitting. The wavenumbers of individual Raman peaks were determined using iterative mathematical methods that minimized the residuals between theoretical and experimental spectra. Figure 1b) presents the pressure-dependent evolution of the wavenumbers corresponding to the most prominent peaks associated with $SiO_4$-related modes, as reported in the literature [28, 36-39]. Notably, distinct inflection points and discontinuities are observed at approximately 5 GPa and 15 GPa.

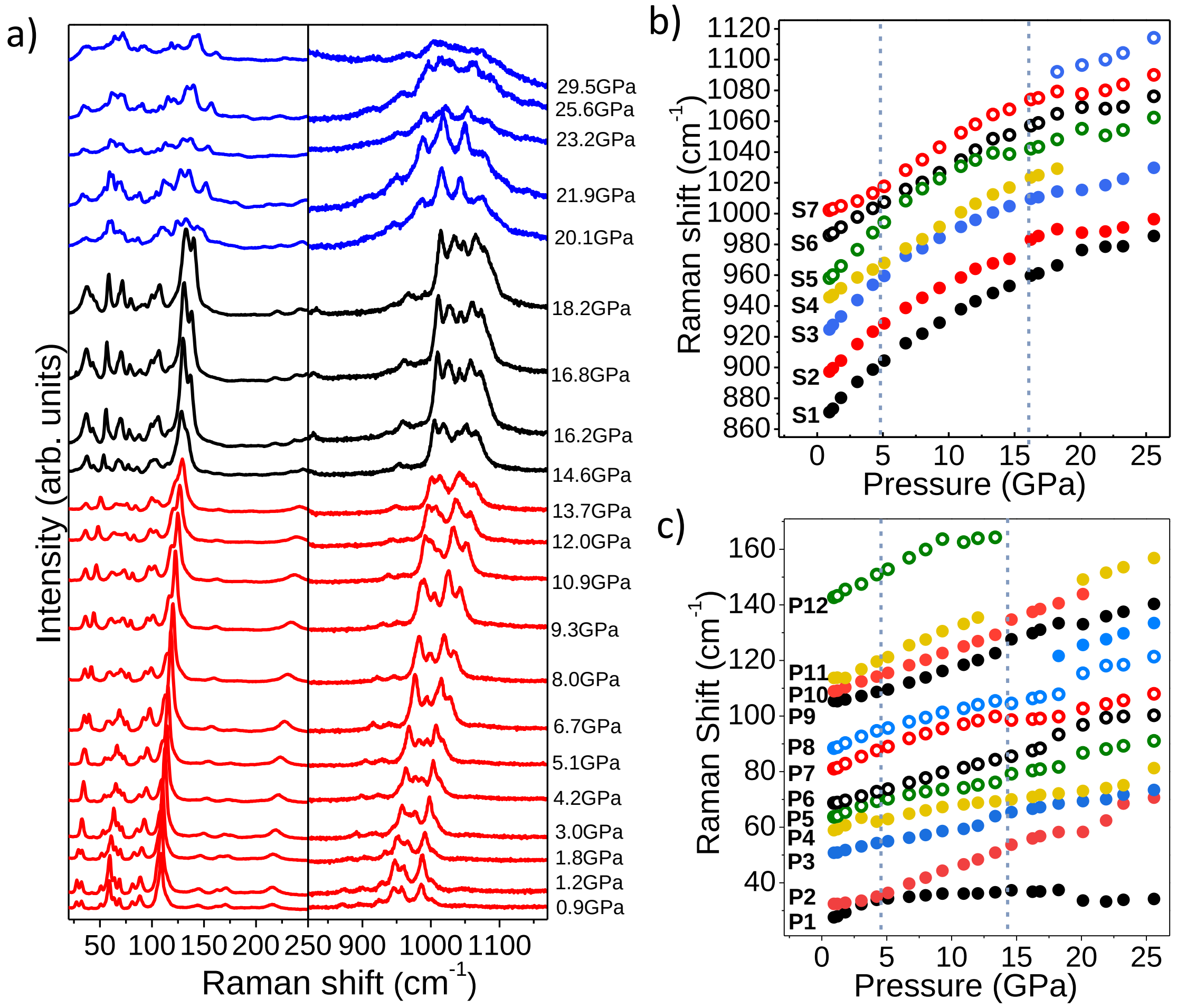

**Figure 1: a)** Raman spectra as a function of high pressures for the A-PS structure. The spectral regions of higher wavenumbers, where the $SiO_4$-related vibrations are located, and lower wavenumbers, corresponding to Pb-O type vibrations, are shown in detail. Pressure evolution of the peaks in the range of $SiO_4$ and Pb-O vibrations, are shown in **b)** and **c)**, respectively. The vertical dashed lines indicate the pressure points of inflection or discontinuity.

To quantify the differences in the evolution of $SiO_4$ vibrational modes between the pressure ranges of ambient to 5 GPa and 5 GPa to 14 GPa, the rates of change of wavenumbers with pressure (*∂ω/∂P*) were determined and are presented on Table 1. For most vibrational modes, significant differences in *∂ω/∂P* values are evident between these two pressure intervals, confirming that the application of hydrostatic pressures exceeding 5 GPa induces structural rearrangements within the $SiO_4$ environments. Additionally, alterations such as jumps, steps, and even splittings in the behavior of the vibrational modes around ~15 GPa suggest more striking reorganizations of silicate arrangements under higher pressure conditions.

**Table 1:**
Pressure variation ($\partial\omega/\partial P$) of the wavenumbers for the high-wavenumber peaks ($SiO_4$ vibrations) and for the low-wavenumber modes (Pb-O vibrations) of A-PS structure, measured in ($cm^{-1}$/GPa).

| $SiO_4$ vibration | | | Pb-O vibration | | |
|---|---|---|---|---|---|
| | Pressure range | | | Pressure range | |
| Peak | P<5GPa | 5GPa<P<13GPa | Peak | P<5 GPa | 5GPa<P<13GPa |
| S1 | 8.5 | 5.3 | P1 | 2.0 | 0.2 |
| S2 | 8.0 | 4.7 | P2 | 0.8 | 1.7 |
| S3 | 8.7 | 4.8 | P3 | 1.1 | 1.0 |
| S4 | 5.5 | 5.5 | P4 | 2.2 | 0.8 |
| S5 | 8.8 | 5.3 | P5 | 1.7 | 0.7 |
| S6 | 5.5 | 4.9 | P6 | 1.3 | 1.3 |
| S7 | 3.2 | 5.7 | P7 | 2.0 | 1.3 |
| | | | P8 | 1.9 | 1.2 |
| | | | P9 | 1.0 | 1.6 |
| | | | P10 | 1.6 | 1.7 |
| | | | P11 | 1.8 | 2.0 |
| | | | P12 | 2.4 | 1.3 |

A similar quantitative analysis was performed to examine the behavior of peaks in the lower-wavenumber region associated with Pb–O vibrations. Using the same spectral fitting method, the pressure-dependent evolution of the main peaks in this region was determined and is presented in Figure 1c). As in the case of $SiO_4$ vibrations, the evolution of most peaks present significant change in slope when pressure exceeds 5 GPa. Table 1 summarizes the $\partial\omega/\partial P$ coefficients for the pressure intervals below and above 5 GPa, enabling a quantitative comparison of slope variations between these two pressure regimes. Figure 1c) also clearly demonstrates that the wavenumbers of the lower-energy vibrational modes undergo significant renormalizations as pressure approaches ~15 GPa, suggesting that the Pb–O arrangements also experience substantial structural alterations under these conditions. In fact, recent reports suggest that significant structural changes outside of the Si-O environments occur in silicate glasses at pressures exceeding 6 GPa. The authors report that in aluminosilicate systems, these pressure values may even cause coordination changes in the counter cation atoms in the structure, resulting in permanent compressions [40].

The analyses of pressure-induced structural evolution using Raman spectroscopy are complemented HP-XRD measurements. Utilizing an X-ray beam energy of 25 keV, the angular range of interest, encompassing the first-order reflection peaks corresponding to the principal planes of the monoclinic A-PS structure, lie in the $4.1° \leqslant 2\theta \leqslant 9.2°$ range. Figure 2a) displays selected diffraction patterns of the A-PS phase measured under increasing hydrostatic pressures, highlighting the high sensitivity of these patterns upon compression up to 24 GPa. Qualitative changes in the X-ray patterns, such as sudden shifts in peak positions and relative intensities, clearly indicate significant structural transformations at pressures above 14 GPa. These structural modifications align with the pressure at which abrupt changes in the evolution of vibrational modes are observed. For instance, the detailed range of diffraction patterns shown in Figure 2b) reveals that for P > 14 GPa, the peaks corresponding to reflections from the $(103)$ and $(30\bar{1})$ planes disappear. Similarly, in another region detailed in Figure 2c), it is evident that at pressures exceeding 15 GPa, the reflection peak from the $(012)$ plane splits, concomitant with the emergence of two new peaks—one to the left of the $(210)$ peak and the other to the left of the $(211)$ peak. As pressure is further increased, the broadening of peak widths strongly suggests a reduction in the degree of long-range order with monoclinic symmetry at pressures above 15 GPa. These findings also align with the high-pressure Raman analyses, which indicate significant reorganizations of both the silicate frameworks and Pb–O bonds as the pressure approaches 15 GPa.

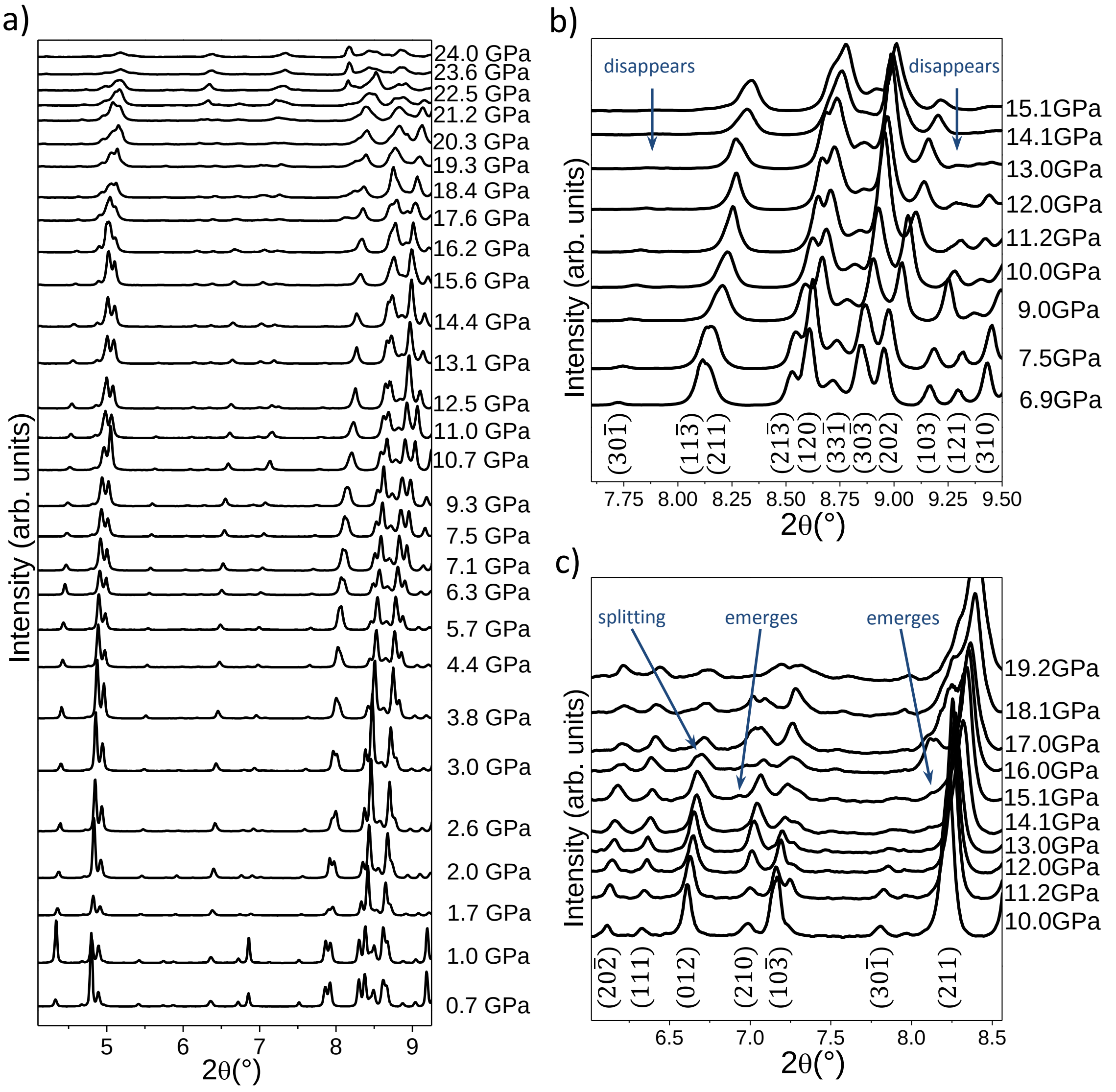

**Figure 2: a)** X-ray diffraction patterns of monoclinic A-PS measured as a function of high pressures. **b)** Reflection peaks from the (103) and $(30\bar{1})$ planes of the alamosite-PS structure disappear for pressures P > 14 GPa. **c)** At P ≥ 15 GPa, the (012) peak splits, and two new peaks emerge to the left of the (210) and (211) peaks.

The quantitative evolution of the A-PS phase was evaluated through structural calculations derived from high-pressure diffraction patterns. Lattice parameters were determined analytically for each pressure point using the least squares method[41]. Figure 3a) depicts the pressure-dependent evolution of the lattice parameters for the monoclinic structure, calculated up to approximately 12 GPa. At higher pressures, the calculations become increasingly imprecise and the resulting lattice parameters are physically unrealistic, indicating the onset of structural destabilization in the alamosite phase.

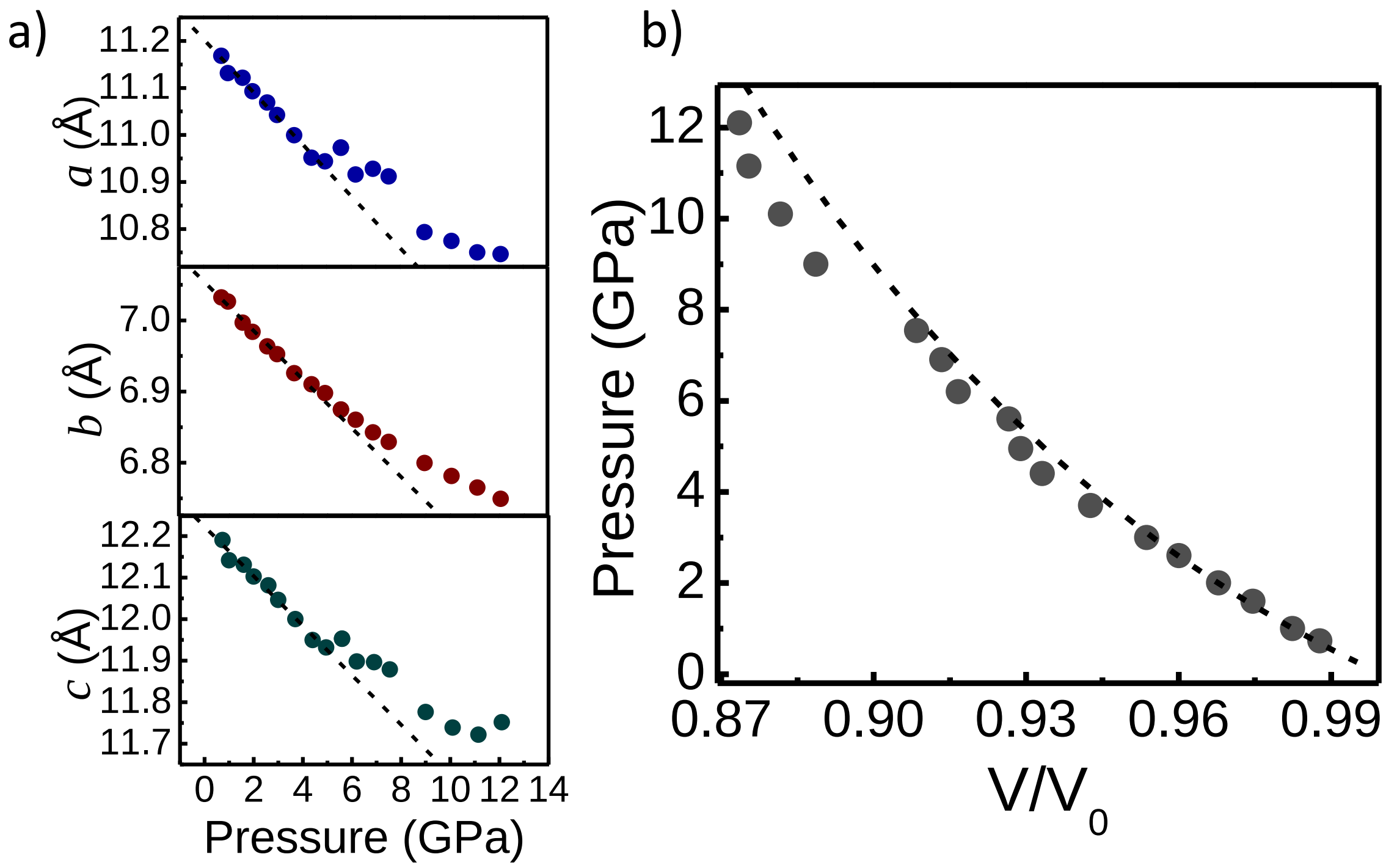

**Figure 3: a)** Lattice parameters of the A-PS (monoclinic) structure as a function of pressure. **b)** Variation of the A-PS structure's volume during compression.

All the lattice parameters exhibit continuous contraction up to 5 GPa, with deviations from linear behavior observed at higher pressures, which align with the analysis of the vibrational modes in Table 1 where *∂ω/∂P* for most of the modes changes upon exceeding 5 GPa, because of rearrangements in the $SiO_4$ and Pb-O environments.

Considering only the data up to 5 GPa, where no changes in the symmetry occur, the axial compressibility factor *K* was calculated for each unit cell axis, yielding the following values: $K_a$ = 4.94 X $10^{-3}$ $GPa^{-1}$ ; $K_b$ = 4.95 X $10^{-3}$ $GPa^{-1}$ e $K_c$ = 4.62 X $10^{-3}$ $GPa^{-1}$. The similarity in compressibility coefficients across all three spatial directions indicates that the compression of the A-PS structure is entirely isotropic within a moderate pressure regime (P < 6 GPa). From these structural calculations, the variation in unit cell volume under hydrostatic pressure ($V/V_0$) ratio was determined. By applying the third-order Birch–Murnaghan isothermal equation of state, the pressure-dependent behavior of the volume was characterized (Figure 3b), expressed as:

$$P(V) = \frac{3B_0}{2}\left[\left(\frac{V_0}{V}\right)^{7/3} - \left(\frac{V_0}{V}\right)^{5/3}\right]\left\{1 + \frac{3}{4}(B_0' - 4)\left[\left(\frac{V_0}{V}\right)^{2/3} - 1\right]\right\}$$

Using this approach, the bulk modulus and its first derivative were calculated as $B_0$ = 52 GPa and $B_0'$ = 10 GPa, respectively. From Figure 3b), it is noteworthy that for P > 6 GPa, the experimental data notably deviates downward from the prediction of the isothermal compression model based on the equation of state. This discrepancy suggests a transition to a distinct compressibility regime, indicative of structural softening. A comparable abrupt compression behavior within this pressure range has been reported in certain aluminum silicates, where it has been attributed to a notable reduction in Si–O bond distances occurring at 6-8GPa[42]. In contrast, some boron silicates exhibited homogenous compression across this pressure range, continuing steadily up to 12 GPa [43]. Given that lead (Pb) is substantially heavier than aluminum (Al), which itself is heavier than boron (B), this observation is consistent with the influence of heavier counter-cations, which tend to reduce the pressure required for phase transitions.

### *3.2. Hexagonal phase (H-PS)*

In Figure 4a) are presented the high-pressure Raman spectra of the metastable hexagonal lead metasilicate structure (H-PS). With increasing pressure, the spectra exhibit peak shifts toward higher wavenumbers, characteristic of mode hardening caused by hydrostatic compression. Additionally, noticeable changes in the Raman profiles occur at specific pressure thresholds. These transitions are evident in the progression from the red spectra to the black spectra and then from the black spectra to the blue spectra. Significant variations in the relative peak intensities are also observed during these transitions between pressure regimes.
Figure 4b) shows the evolution of vibrations related to $SiO_4$ units derived from spectral fitting, highlighting inflections and/or discontinuities around 5 GPa and ~14 GPa, as indicated by dashed lines. $\partial\omega/\partial P$ of these modes were calculated for two distinct pressure ranges: P < 5 GPa and 5 GPa < P < 13 GPa. The values are summarized in Table 2, where notable differences in $\partial\omega/\partial P$ for most vibrational modes are evident across these intervals. Here again, such deviations from a continuous trend suggest changes in the short-range symmetries of the structures formed by $SiO_4$ units, which proves that pressures near 5 GPa are capable to induce structural rearrangements in the $SiO_4$ environments also in H-PS phase. At pressures above 13 GPa, further spectral changes, including abrupt jumps, steps, and splittings, point to additional reorganizations of the silicate frameworks.

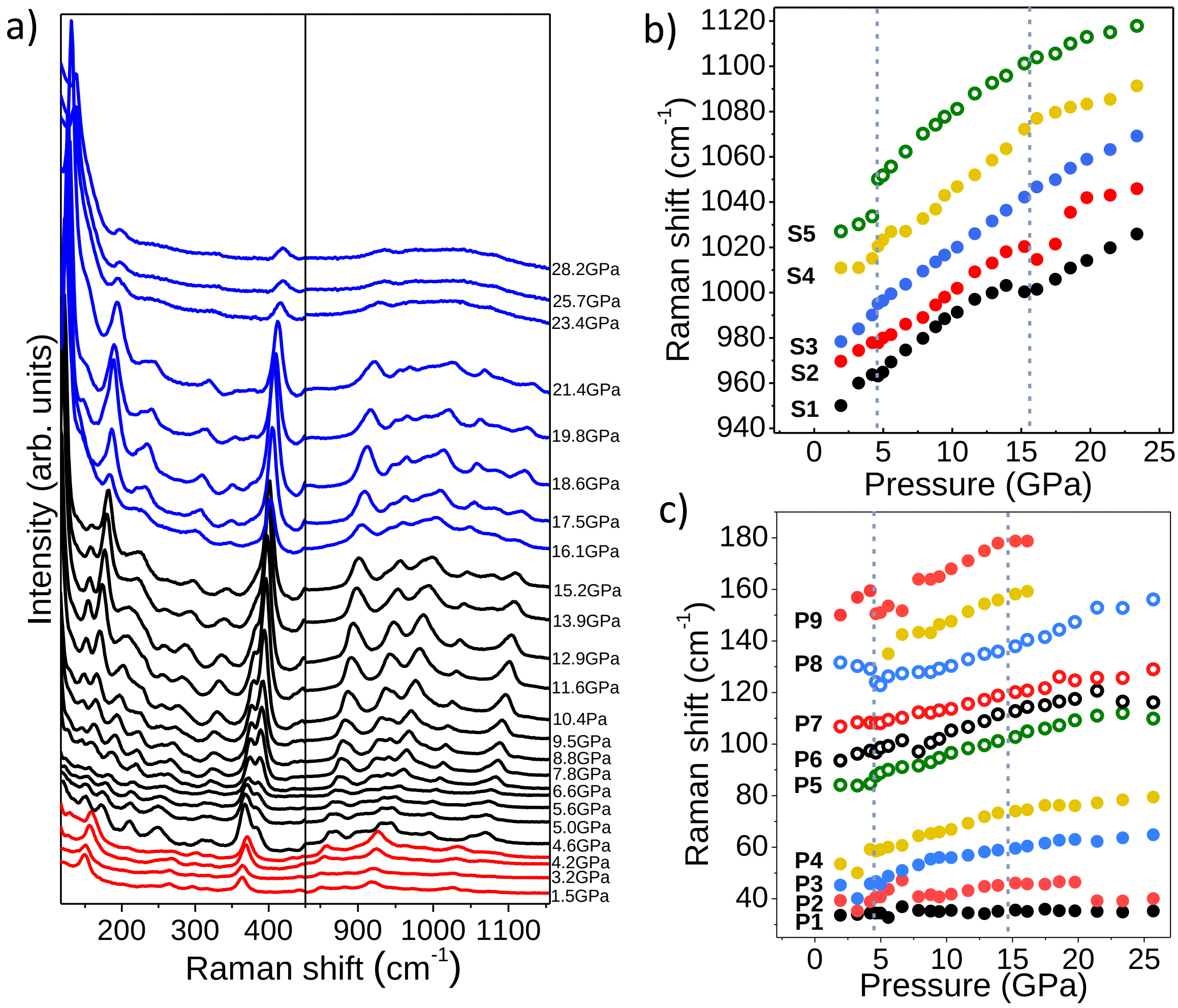

**Figure 4: a)** Raman spectra as a function of high pressure for the H-PS structure. Pressure evolution of the peaks in the range of **b)** $SiO_4$ vibrations and **c)** Pb-O vibrations of the H-PS structure

Figure 4c) illustrates the pressure dependence of lower-wavenumber modes associated with Pb-O vibrations, which also exhibit significant changes in their behavior as pressures exceed ~5 GPa. Additionally, these peaks undergo further renormalization upon reaching pressures of 15–16 GPa, with some modes disappearing entirely. This behavior parallels that of the $SiO_4$ modes, providing further evidence of substantial structural changes involving both the Pb–O and $SiO_4$ environments. $\partial\omega/\partial P$ for the peaks within two pressure intervals P < 5 GPa and 5 GPa < P < 13 GPa are summarized in Table 2.

**Table 2:**
Pressure variation $\partial\omega/\partial P$ of the wavenumbers of the high-wavenumber modes ($SiO_4$ vibrations) and for low-wavenumber modes (Pb-O vibrations) for H-PS structure ($cm^{-1}$/GPa).

| $SiO_4$ vibration | | | Pb-O vibration | | |
|---|---|---|---|---|---|
| | Pressure range | | | Pressure range | |
| Peak | P<5GPa | 5GPa<P<13GPa | Peak | P<5 GPa | 5GPa<P<13GPa |
| S1 | 6.1 | 4.4 | P1 | 0.2 | 0.1 |
| S2 | 3.6 | 4.4 | P2 | -3.2 | 3.3 |
| S3 | 5.2 | 4.5 | P3 | -4.1 | 1.4 |
| S4 | 1.8 | 4.9 | P4 | -2.7 | 1.6 |
| S5 | 2.9 | 3.2 | P5 | 0.2 | 1.4 |
| | | | P6 | 2.0 | 1.4 |
| | | | P7 | 0.7 | 1.1 |
| | | | P8 | -1.1 | 1.3 |
| | | | P9 | 4.2 | disappears |

A comparison of the data in Table 1 and Table 2 reveals that the $\partial\omega/\partial P$ coefficients associated with the $SiO_4$-related modes are generally higher in the A-PS phase than in the H-PS structure, particularly at pressures below 5 GPa. This observation suggests that the $SiO_4$ bonding environments in the stable A-PS phase are more responsive to pressure effects compared to those in the H-PS structure. Hypotheses derived from studies on quartz propose the presence of distinct compression mechanisms throughout this entire pressure range. For pressures P < 6 GPa, the stress exerted on the material is accommodated by twisting of the Si-O-Si bonds and rotations of the $SiO_4$ tetrahedra. A direct compression of the $SiO_4$ tetrahedra may occur only at pressures exceeding 6 GPa[44]. Based on this model, the differences in $\partial\omega/\partial P$ of $SiO_4$ modes between the two structures at P < 5 GPa suggest that such pre-compression effects (bond torsion and tetrahedral rotation) are more pronounced in A-PS than in H-PS. Conversely, the data pertaining to Pb–O vibrations do not provide conclusive evidence to ascertain in which crystalline polymorph of $PbSiO_3$ this type of bond is more significantly affected by compression.

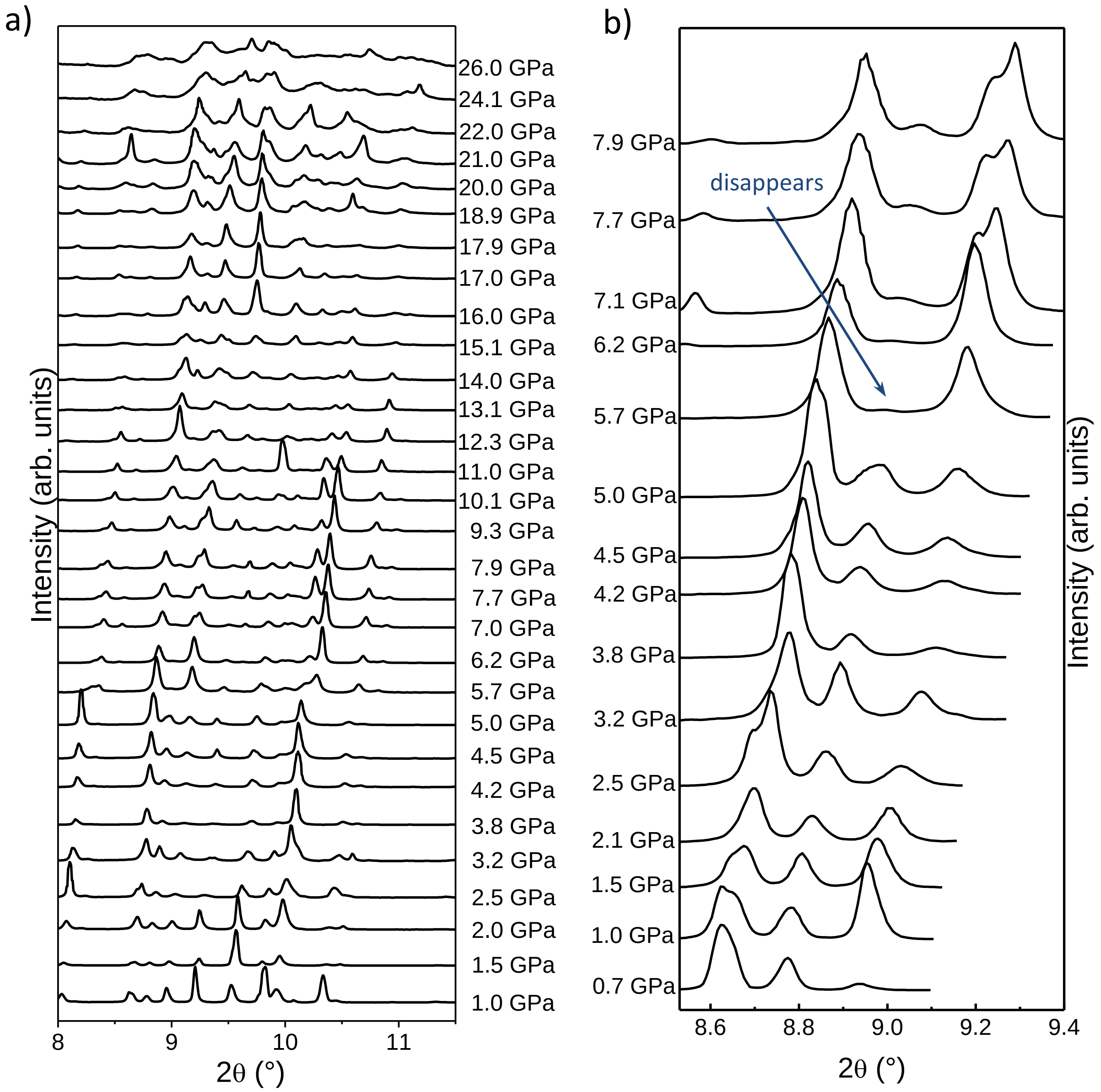

**Figure 5: a)** X-ray diffraction patterns of the H-PS phase as a function of high pressures. **b)** Selected region in moderate pressure regime.

The compression of H-PS structure was also analyzed using diffraction patterns recorded at increasing hydrostatic pressure, presented in Figure 5a). For pressures above 14 GPa, a pronounced broadening of the diffraction lines is observed, consistent with a reduction in the degree of the crystalline symmetry. A more detailed analysis of the H-PS structure was performed in selected 2θ region illustrated in Figure 5b). Up to ~5 GPa, the diffraction peaks exhibit continuous shifts toward higher angles, confirming the structural stability of the H-PS phase within this pressure range. This observation aligns with the monotonic behavior of the $SiO_4$ and Pb–O vibrational modes shown in Figures 4b) and 4c), respectively. For P > 5 GPa, the relative intensity of the central peak sharply declines (blue arrow) until it disappears. The suppression of diffraction peaks is generally indicative of structural alterations involving changes in symmetry. Significantly, this sudden modulation in diffraction patterns coincides with the pressure at which discontinuities are observed in the evolution of Raman modes. The fact that abrupt changes in both diffraction and Raman signal behavior occur within the same pressure range as the phase transition in the A-PS structure suggest that the microstructural transformations in both polymorphs are driven by analogous mechanisms. Mirroring the methodology previously applied to the A-PS crystal structure, a more comprehensive quantitative analysis of the H-PS structure was conducted, focusing on moderate pressure regimes. In Figure 5b) the diffraction peaks in the 8.5 < 2θ < 9.1 interval correspond to reflections from the $(100)$, $(002)$, and $(101)$ crystalographic planes. The lattice parameters at each pressure were determined analytically from these reflections using least-squares calculations[41]. Figure 6a) presents the evolution of the lattice parameters as a function of pressure, which was subsequently used to calculate the compression of the H-PS unit cell volume, depicted in Figure 6b). The bulk modulus of the H-PS structure was determined using the third-order Birch–Murnaghan isothermal equation of state, yielding a value of $B_0$ = 48 GPa and its first derivative $B_0'$ = 6 GPa. The modest bulk modulus suggests that the H-PS is highly susceptible to pressure-induced modifications. Furthermore, it is worth noting that the value of $B_0$ for the H-PS structure is lower than that of the A-PS structure ($B_0$ = 52 GPa). This difference is consistent with the metastable nature of the H-PS structure regard to temperature, reflecting its lower energy state in comparison to the stable A-PS phase.

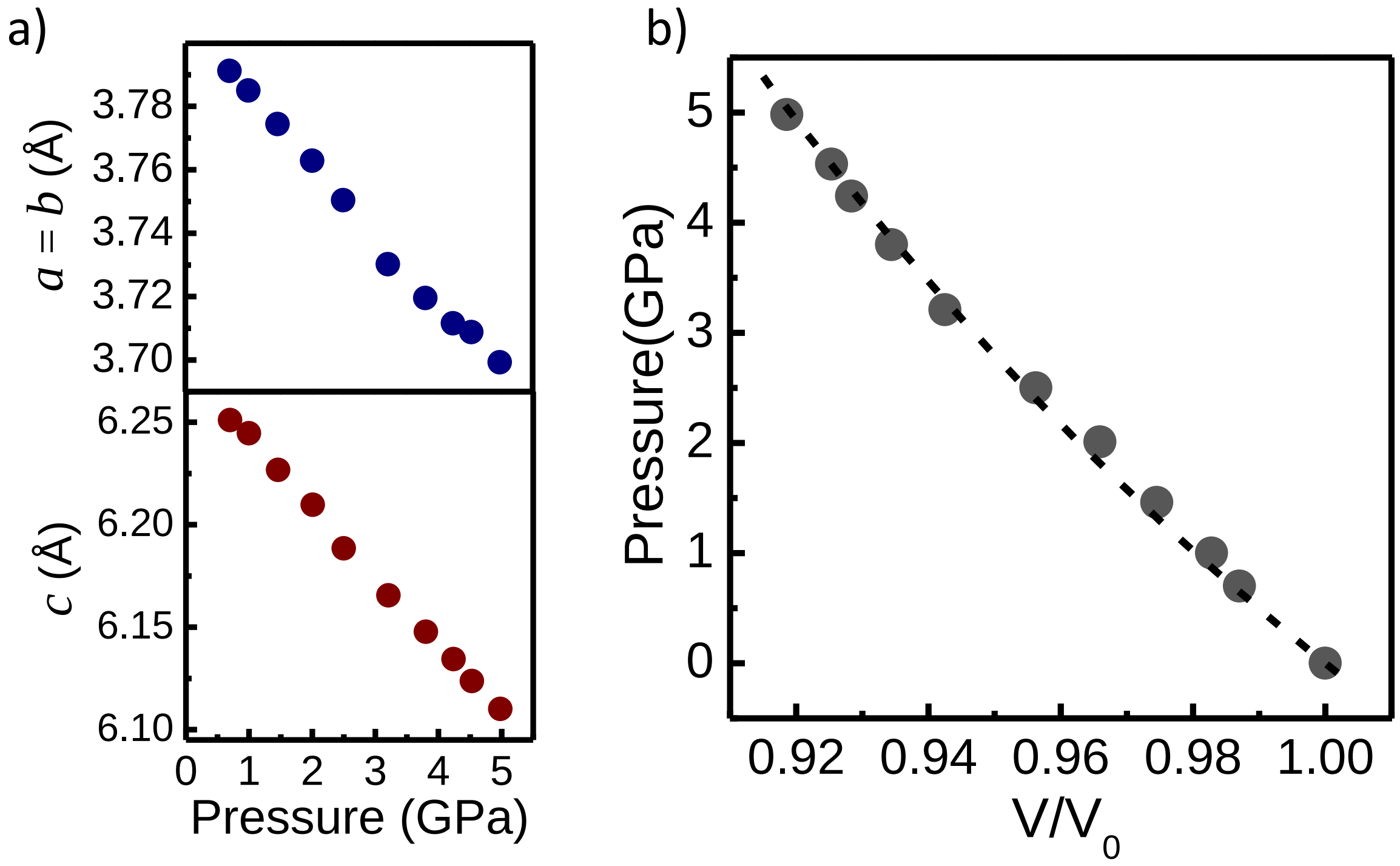

**Figure 6: a)** Evolution of the lattice parameters of H-PS with pressure. **b)** Relative volume of the unit cell as a function of pressure and fitting with the third-order Birch–Murnaghan isothermal equation of state (dashed line).

### *3.3. Insights into the lower symmetry phase (L-PS)*

In addition to the monoclinic and hexagonal structures, the effects of hydrostatic pressure were also examined for another metastable phase of PS with low symmetry, denoted as L-PS, whose crystalline symmetry are undetermined. Figure 7a) displays the high-pressure diffraction patterns of the L-PS phase, whereas Figure 7b) ilustrates the angular positions of the principal diffraction peaks, derived from fitting the diffractograms at each pressure point. The progression of the peaks exhibits a nearly continuous trend up to 8 GPa. Above this pressure, alterations such as inflections, peak splitting, and the disappearance of some peaks become apparent, signifying substantial structural modifications. Notably, from 10 GPa onward, the majority of the diffraction peaks display a broadening in their profiles, indicative a significant rise in the degree of disorder and a corresponding reduction in long-range crystalline order. These results demonstrate that the low-symmetry L-PS structure exhibits greater resilience to applied pressure at moderate regime compared to the A-PS and H-PS structures, which undergo structural transformations at approximately 5 GPa.

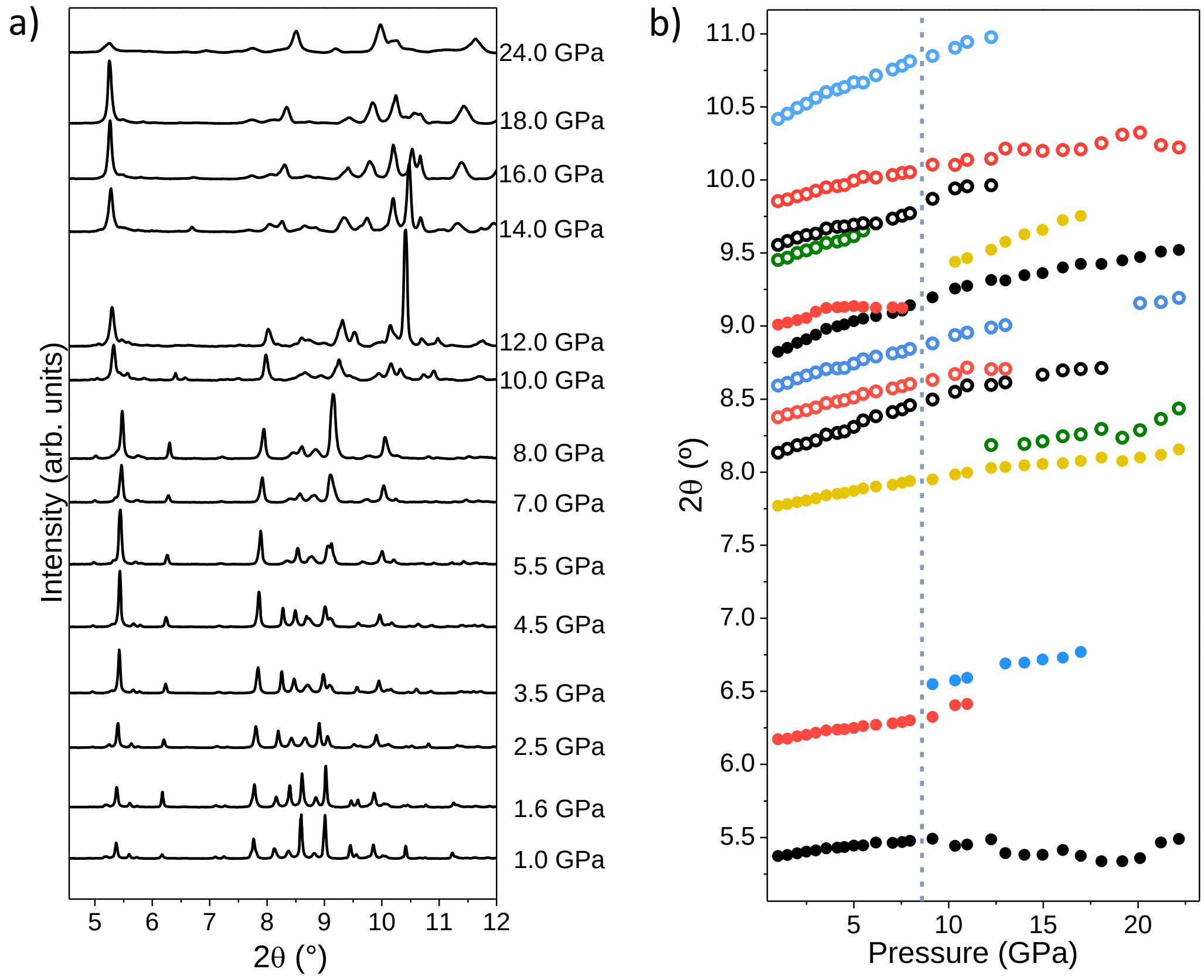

**Figure 7: a)** Selected X-ray diffraction patterns of the L-PS phase as a function of high pressures. **b)** Evolution of the main diffraction peaks with pressure.

## 4. Conclusions

The analyses presented here demonstrate that significant structural changes can be induced in lead silicate crystalline polymorphs under moderate compression environment. The combined use of high-pressure Raman spectroscopy and high-pressure synchrotron x-ray diffraction proved to be an effective approach for detecting structural transitions in the silicate environment within moderate pressure regimes ($P < 10$ GPa), which contrasts with previous studies in the literature [45-47], where other in situ analytical techniques were less effective in identifying structural transformations within this pressure range. Diffraction analyses of the stable monoclinic phase of lead metasilicate, alamosite (A-PS), and a metastable hexagonal phase (H-PS) revealed distinct pressure-dependent structural responses for P < 5 GPa and P > 5 GPa. Moreover, inflection points and/or discontinuities in the evolution of the vibrational modes indicate that these transitions are accompanied by rearrangements of the $SiO_4$ units. This observation implies that pressures at this scale can induce structural even within the fundamental silicate units, independent of whether they occur in the lower-energy hexagonal phase (H-PS) or the higher-energy monoclinic structure (A-PS). Additionally, the phase transition mechanism entails significant alterations in the chemical environment of counter-cations, as evidenced by variations in the pressure evolution of Pb-O vibrational modes. It was further observed that $SiO_4$ vibrations exhibit a greater sensitivity to pressure in the A-PS phase compared to the H-PS structure, suggesting distinct microstructural deformation mechanisms between the two phases. Coupled with the differing bulk moduli of these phases, these findings underscore the complexity of the crystallization process under pressure, with each crystal–crystal phase transformation likely proceeding through a unique compression pathway. Furthermore, we analyzed an additional metastable, low-symmetry phase of lead metasilicate (L-PS). This phase undergoes continuous compression up to approximately 8 GPa, beyond which an increase in structural disorder is observed, accompanied by a decline in long-range crystallinity. Notably, this transition occurs at a lower pressure relative to the other phases, in which a significant rise in structural disorder is only detected at pressures exceeding 14 GPa. These findings not only provide crucial insights for extending this understanding to other silicate systems, but also highlight the potential to influence the multi-stage crystallization process of silicates through the application of pressure, presenting an alternative to conventional temperature-driven methods.

**Acknowledgments**

We acknowledge Prof. Michel Venet Zambrano for his invaluable assistance with the structural calculations. This study was financed, in part, by the São Paulo Research Foundation (FAPESP), Brasil. Process Numbers #2021/13974-0 (ADR), #2019/12383-8 (TRC), #2017/11868-2 (RBP), #2013/07793-6 (PSP); and the National Council for Scientific and Technological Development (CNPq) process 310819/2023-7. We also wish to thank the Brazilian Synchrotron Light Laboratory (LNLS/CNPEM) - EMA beamline and LCTE (Proposal EMA-20210198).